\documentclass[10pt,letterpaper]{article}
\usepackage{opex3}
\usepackage{subfig}
\usepackage{graphicx}
\usepackage{float}
\usepackage{amssymb, amsmath}
\begin{document}

\title{Non-classical interference in integrated 3D multiports}

\author{Thomas Meany,$^{1} $Michael Delanty,$^{1}$ Simon Gross,$^{1}$  Graham D. Marshall,$^{1,2}$ M. J. Steel,$^{1}$ and Michael J. Withford$^{1}$}

\address{$ ^{1}$ Centre for Ultra-high bandwidth Devices for Optical Systems (CUDOS),\\ MQ Photonics Research Centre, Department of Physics and Astronomy, \\Macquarie University, North Ryde, 2109 NSW, Australia \\ $^{2}$ Centre for Quantum Photonics, \\H. H. Wills Physics Laboratory and Department of Electrical and Electronic Engineering,\\ University of Bristol, Merchant Venturers Building, Woodland Road, Bristol, BS8 1UB, UK}

\email{thomas.meany@mq.edu.au} 



\begin{abstract*} 
We demonstrate three and four input multiports in a three dimensional glass platform, fabricated using the femtosecond laser direct-write technique. Hong-Ou-Mandel (HOM) interference is observed and a full quantum characterisation is performed, obtaining two photon correlation matrices for all combinations of input and output ports.  For the three-port case, the quantum visibilities are accurately predicted solely from measurement of the classical coupling ratios. 
\end{abstract*}



\section{Introduction}
Optical multiports provide the ability to produce an arbitrary unitary transformation of a set of optical modes~\cite{Nportunitary}. Multiports are therefore an extremely important component in optical quantum information processing (QIP) and state preparation~\cite{Whiteentangledstate,entangedmultiport}, where producing entangled states is a key motivator for such devices~\cite{Szameit_multiport_Cleo2012, BroughamEPJ}. Multiports and their applications in quantum optics were originally explored using bulk optic beamsplitters and phase shifters~\cite{Tritterbulkoptic}, for which non-classical interference was observed in both three- and four-port splitters. Following from this work a study of a three-port fiber coupler was undertaken~\cite{Tritterfibre}. However neither of these approaches provide the scalability and stability requirements needed for large scale optical quantum circuits.  The extreme difficulty of working with large numbers bulk optical elements and fibre components has limited the development  of these devices. 

The challenges of scaling and stability in various  quantum optical applications have been the main drivers for the recent explosion of interest in integrated quantum devices, such as integrated waveguide circuits~\cite{Politi,zhangGaN}, waveguide array quantum walks~\cite{Peretsqw,Brombergqw,PeruzzoQW} and single photon sources~\cite{PPLNentanged,LPOR}. Desirable qualities  such as reconfigurability have been demonstrated~\cite{Reconfigurable}, notably in lithium niobate-based devices which allow for telecommunications wavelength photon manipulation at GHz speeds~\cite{LiNtelecomcontrol}. Miniaturisation using silicon-on-insulator nanowire based photonic circuits have been shown~\cite{Siwiremachzehnder} and there is also great promise to incorporate these platforms with existing on chip detectors~\cite{Onchipdetectors}. In addition the integration of photon pair sources and circuits have been achieved using a nonlinear waveguide array fabricated in lithium niobate~\cite{quantumwalknonlinear}.  The advantages of integrated components have been applied to the construction of integrated quantum multiports. Peruzzo~\emph{et al.} have demonstrated multiport functionality with an on-chip 4-port multimode interference device~\cite{MMI} in the silica-on-silicon platform.

The majority of these demonstrations have relied on mature planar platforms, but there are limitations associated with the inherently two dimensional (2D) nature of planar quantum circuits. For instance in the realm of optical quantum simulation~\cite{quantumsimulators}, there are interesting Hamiltonians which can only be effectively simulated using waveguides in three dimensional (3D) arrangements~\cite{superradiance}.  Moreover, as circuits become more complicated, both the circuit area and the proportion of ``neutral'' elements, whose only purpose is to provide transparent waveguide crossings, will become significant. Hence techniques which allow the fabrication of structures in three dimensions have some specific and unique possibilities. Such a technique is the femtosecond laser direct-write technique (FLDW) which allows for fabrication of waveguides in glass subsrates in 3D. This technique has been used to produce a number of devices showing
quantum operation including directional couplers~\cite{Grahamoptex} (evanescently-coupled waveguides which act as beamsplitters), a 6-port  continuous quantum walk on an elliptical array~\cite{OwensQW}, a discrete quantum walk on a line using entangled photons~\cite{SansoniQW} and investigations of quantum interference in 2D arrays~\cite{Szameit2Darrays}. This has demonstrated the potential of the FLDW technique to fabricate useful structures for quantum applications.

In this paper we implement a 3-port and 4-port device using a 3D FLDW technique. Each device highlights a different advantage of the 3D geometry. The 3-port is implemented as a single coupling region that replaces three 2-port devices~\cite{Nportunitary}. The 4-port is composed of four 2-port directional couplers, but avoids a crossing that would require an additional coupler in 2D. Taken together these kinds of structures could reduce the complexity and sensitivity of higher order concatenated multiport based circuits. This fact is established by our observation of 2 photon interference in both structures.

\section{Implementing multiports}
The fundamental idealised multiport device is the familiar beamsplitter which implements a desired $2\times 2$ unitary operation.  This can be generalised to a multiport operation on \emph{N} input modes $\hat{a}_N$ into \emph{N} output modes $\hat{b}_N$ described by the unitary $N\times N$ matrix~\cite{Tritterbulkoptic} 
\begin{eqnarray}
\begin{bmatrix}
 \hat{b}_i \\
 \hat{b}_j\\
\vdots \\
\hat{b}_N
  \end{bmatrix} =
 \begin{bmatrix}
  U_{1,1} & U_{1,2} & \cdots & U_{1,N} \\
  U_{2,1} & U_{2,2} & \cdots & U_{2,N} \\
  \vdots  & \vdots  & \ddots & \vdots  \\
  U_{N,1} & U_{N,2} & \cdots & U_{N,N}
 \end{bmatrix}
 \begin{bmatrix}
 \hat{a}_i \\
 \hat{a}_j\\
\vdots \\
 \hat{a}_N
  \end{bmatrix}   .
\end{eqnarray}
The integrated photonics analog of the basic 2-port beamsplitter is the directional coupler, in which two waveguides are brought into close proximity so that their evanescent fields can overlap and exchange energy~\cite{Politi,SNYDER_72}.  The degree of coupling is determined by the waveguide separation and interaction length  which can be adjusted to produce directional couplers of any reflectivity.  Using the well-known decomposition of the $N\times N$ unitary into a concatenation of beamsplitters and phase shifters~\cite{Nportunitary}, it is possible to construct an arbitrary  multiport in two dimensions in integrated optics by designing a suitable arrangement of directional couplers and phase delays. However as the number of directional couplers scales with $N^2$, at large N, such a device is prohibitive to build due to fabrication errors.

 Previous investigations have looked at cases of bulk optic beamsplitters~\cite{Tritterbulkoptic} , fibre optic directional couplers~\cite{Tritterfibre}, and multimode interference devices~\cite{MMI}, to produce multiports In the next section we apply these techniques to the specific cases of a three and four port beamsplitter based on an arrangement of coupled waveguides

\subsection{Mode evolution in the three port (tritter)}

\begin{figure}[h]
  \centering
  \subfloat[]{\label{fig:Tritterideal}\includegraphics[width=0.25\textwidth]{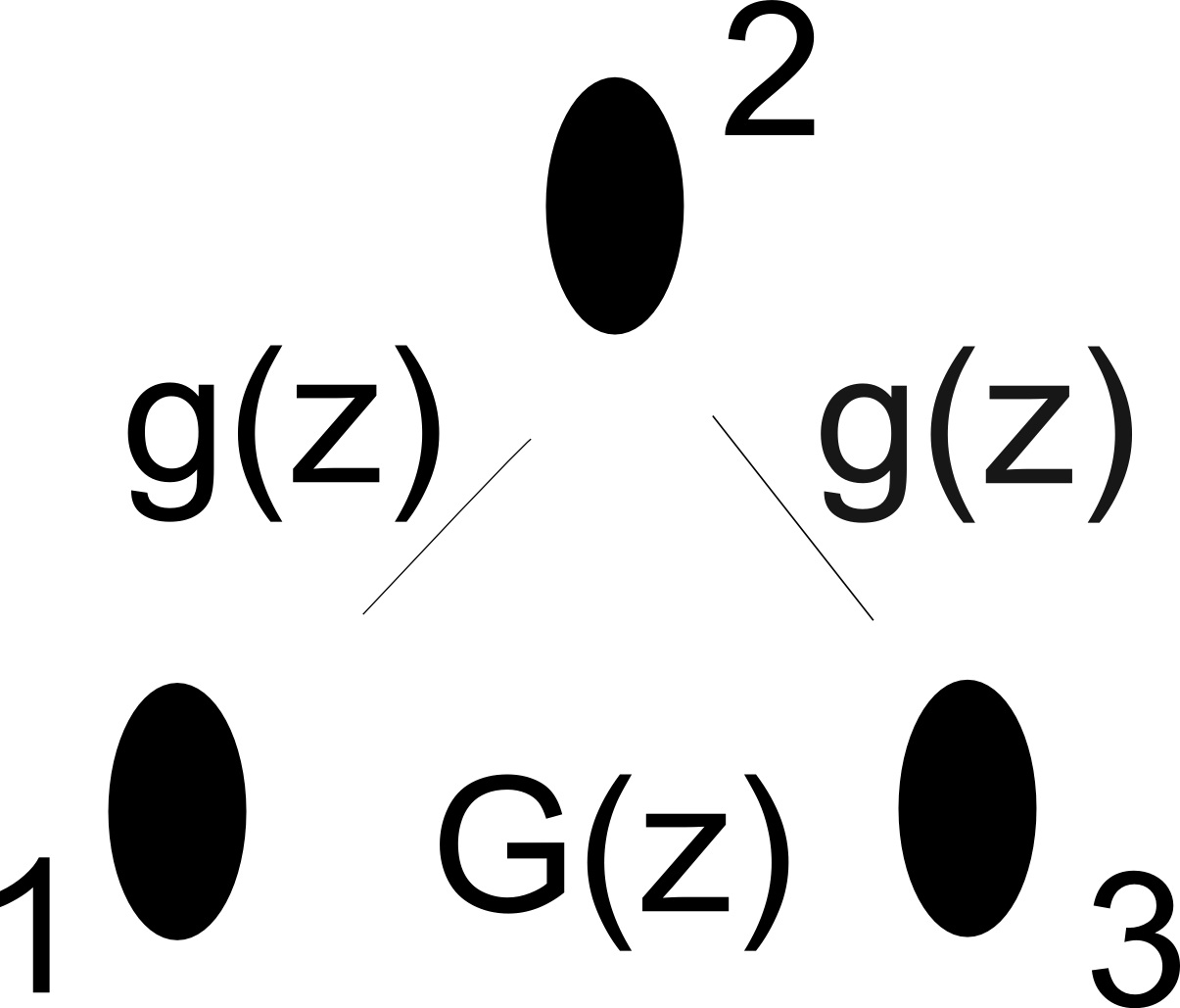}} 
  \subfloat[]{\label{fig:Tritter}\includegraphics[width=0.38\textwidth]{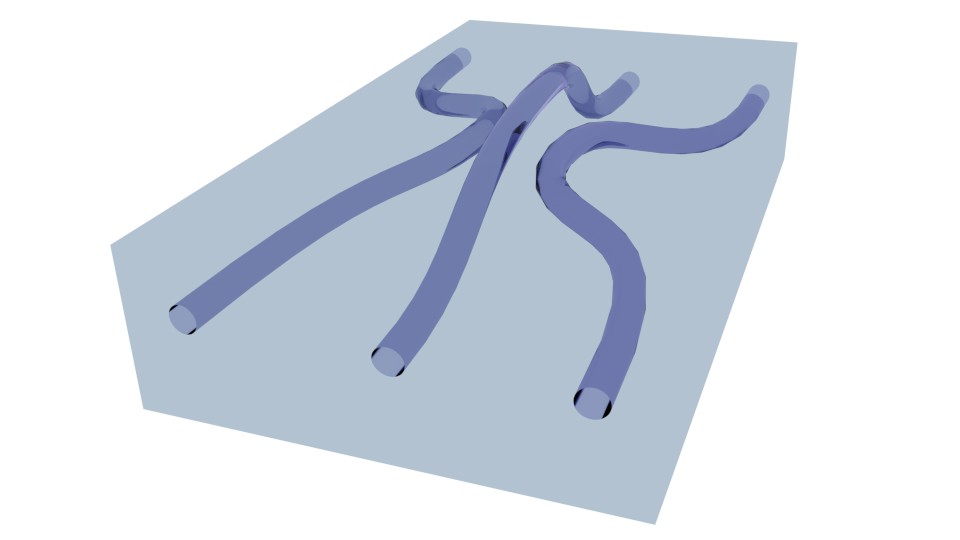}}
 \subfloat[]{\label{fig:4port}\includegraphics[width=0.38\textwidth]{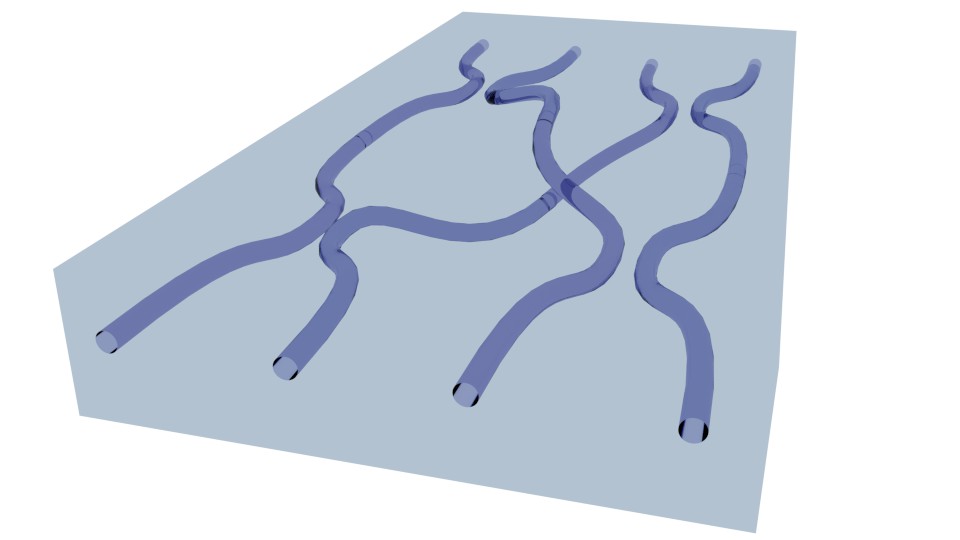}} 
 \vspace {-1mm}
\caption{
(a) Idealised coupling region of a tritter device. Writing considerations induce an effective isosceles geometry with two coupling strengths $g(z)$ and $G(z)$.
(b) A schematic of a 3D implementation of a tritter, where three waveguides taper from a planar arrangement to a triangular interaction region and taper back to a planar array and (c) the layout of the four directional couplers and the waveguide avoided crossing in the 4-port device}  \vspace {-1mm}
  \label{fig:tritter4port}
\end{figure}

In three dimensions a $3\times 3$ multiport or tritter can be constructed from a single coupling element involving three waveguides (see Fig.~\ref{fig:Tritterideal}). This single element can replace up to three directional couplers and phase shifters~\cite{Nportunitary}. The coupling coefficient $G(z)$ describes coupling between waveguides $1$ and $3$, and $g(z)$ describes the coupling between waveguides $1$ and $2$ and waveguides $2$ and $3$.  In general, all three of the couplings could have different values, but, as will be discussed later, the 'isosceles'� geometry considered here is appropriate as a result of the laser writing process we used. 

 \vspace {-1mm}

 Such a structure is described by a Hamiltonian of the form
\begin{equation}
\label{GeneralHam}
\hat{H} = \hbar v_p \sum^{3}_{n,m=1} C_{n,m} \hat{a}^{\dagger}_n \hat{a}_m,
\end{equation}
where the coupling matrix is
\begin{equation}
C =
\begin{bmatrix}
 \beta & g  & G \\
 g & \beta  & g \\
 G & g  & \beta 
  \end{bmatrix}.
\end{equation}
The propagation constant $\beta=\omega/v_p$ of the three waveguide modes is assumed to be common ($v_p$ is the phase velocity and $\omega/2\pi$ the optical frequency).  Explicitly including the common phase velocity $v_p$ in $\hat{H}$ means that operator evolution with distance $z$ rather than time $t$ follows naturally.

After propagation through the interaction region, photons have encountered the mode transformation 
\begin{eqnarray}
\mathbf{\hat{b}} = U \mathbf{\hat{a}}
\label{Ueqn}
\end{eqnarray}
where the input and output mode vectors are $\mathbf{\hat{a}} = (\hat{a}_1, \hat{a}_2, \hat{a}_3)$ and  $\mathbf{\hat{b}} = (\hat{b}_1, \hat{b}_2, \hat{b}_3)$, and the transfer matrix $U$ is the solution of Heisenberg's equation, $d \mathbf{\hat{a}}/dz   = -i C (z) \mathbf{\hat{a}} $. The transfer matrix $U$ completely determines the operation of the device, in particular the classical output intensities and quantum interference between photons in the device.

\subsection{Mode evolution in the four-port}

In contrast to the single element tritter the four port (depicted in Fig.~\ref{fig:4port}) is made up of four directional couplers. Here we have used the 3D advantage to cross two waveguides without interaction, reducing the number of required directional couplers by one.
As the device is made up of four discrete elements, the device can be modelled as a product of beam splitter operators. Assuming each directional coupler is identical, the operator corresponding to the four port transfer matrix is,
\begin{eqnarray}
\hat{U} &=& \hat{B}_{2,4} (\eta) \hat{B}_{1,3} (\eta) \hat{P}_2(\phi) \hat{P}_3(\phi) \hat{B}_{3,4} (\eta)\hat{B}_{1,2} (\eta) \label{FourPortU}
\end{eqnarray}
where the directional couplers have the beam splitter action,
\begin{eqnarray}
\begin{bmatrix}
 \hat{b}_i \\
 \hat{b}_j
  \end{bmatrix} = \hat{B}_{i,j} (\eta) \begin{bmatrix}
 \hat{a}_i \\
 \hat{a}_j
  \end{bmatrix}   =
  \begin{bmatrix}
 \sqrt{\eta} & -i \sqrt{1-\eta} \\
  -i\sqrt{1-\eta} & \sqrt{\eta}
  \end{bmatrix}
\begin{bmatrix}
 \hat{a}_i \\
 \hat{a}_j
  \end{bmatrix}  ,
  \label{bs}
\end{eqnarray}
and $\hat{P}_j(\phi)$ describes the phase difference between the inner crossing arms and the outer ``straight-through'' arms (See Fig.~\ref{fig:4port}) where
 $\hat{P}_j(\phi) \hat{a}_j = e^{-i \phi} \hat{a}_j$. Using these relations we find the transfer matrix corresponding to the four port operator (\ref{FourPortU}) is
\begin{eqnarray}
U=
\begin{pmatrix}
 \eta  & -i e^{i \phi } \sqrt{(1-\eta ) \eta } & -i \sqrt{(1-\eta ) \eta } & e^{i \phi } (\eta -1) \\
 -i \sqrt{(1-\eta ) \eta } & e^{i \phi } \eta  & \eta-1  & -i e^{i \phi } \sqrt{(1-\eta ) \eta } \\
 -i e^{i \phi } \sqrt{(1-\eta ) \eta } & \eta-1  & e^{i \phi } \eta  & -i \sqrt{(1-\eta ) \eta } \\
 e^{i \phi } (\eta-1 ) & -i \sqrt{(1-\eta ) \eta } & -i e^{i \phi } \sqrt{(1-\eta ) \eta } & \eta 
  \end{pmatrix}   .
\label{UMatrixFourPort}
\end{eqnarray}

\section{Fabrication methods and design}

The devices were fabricated using the femtosecond laser direct-write (FLDW) technique which employs a tightly focused femtosecond laser to inscribe a localized refractive index change in glass. By translating the glass sample in the (\emph{x, y, z}) directions with respect to the incident laser, arbitrary 3D regions of net-positive refractive index change (which act as waveguides) can be produced inside the host glass. The laser used to fabricate the structures is a titanium sapphire oscillator (Femtolasers GmbH, FEMTOSOURCE XL 500, 800~nm centre wavelength, $<50$~fs pulse duration) with 5.1~MHz repetition rate. A 100X oil immersion objective (Zeiss N-Achroplan, \emph{NA}~=~1.25) was used to focus the laser inside boro-aluminosilicate (Eagle 2000).  The sample translation was completed using Aerotech high precision motion control stages with 10~nm precision at writing speeds of 2000~mm/min. The waveguides were written using pulse energies of $28~\mu$J which created waveguides suitable for single mode operation at 800~nm. The combination of low pulse energy, high translation speed and high repetition rate created waveguides which were at the at the very onset of cumulative heating modification mechanism~\cite{EatonCOMHeating} (implying that pulses arrive before the characteristic thermal diffusion time of the material). This results in smooth refractive index change and short device production time, allowing the fabrication of complex photonic circuits. The refractive index change is ellipsoidal in nature and  it has been shown that laser written waveguides display non-isotropic coupling due to waveguide asymmetry~\cite{Szameit_nonisotropic_coupling}. Since the tritter is a triangular arrangement of modes (as shown in Fig.~\ref{fig:Tritterideal}) there are two effective coupling ratios, $g(z) \not= G(z)$.
 In both cases the waveguides are initially separated, in a planar array, by 127~$\mu$m before tapering an interaction region, where evanescent coupling occurs with typical spacings of 15~$\mu$m and subsequently tapering out to 127~$\mu$m planar spacing.This is best illustrated in Fig.~\ref{fig:tritter4port}, which shows the waveguides tapering towards interaction regions and returning to the original spacign at the output of the device. We set out to fabricate a tritter with equal splitting between output ports of 33/33/33 and a 4-port with  equal splitting between output ports of 25/25/25/25. Hence we fabricated a range of devices with different length interaction regions targeting the ideal coupling ratios. We performed a classical characterisation to determine which structures best matched our intended constraints. 

\section{Classical characterisation \label{ClassicalSection}}
In order to determine the coupling ratios in the tritter and  splitting ratios for the four port a classical characterisation was performed.  If classical light of intensity $M$ is injected into port $j$, the output intensity $N$ at port $k$ can be modelled as, 
\begin{eqnarray}
N^k_j &=& \epsilon^{in}_j \epsilon^{out}_k M |U_{j,k}|^2,
\end{eqnarray}
where the input and output losses at port $l$ are characterized by the pre-factors, $0<\epsilon^{in/out}_l<1$. By forming the following ratios,
\begin{eqnarray}
F^{k,s}_{j,r}&\equiv &\frac{N^k_j N^s_r}{N^k_r N^s_j} = \frac{|U_{j,k}|^2 |U_{r,s}|^2}{|U_{r,k}|^2 |U_{j,s}|^2}  ,
\label{Fractions}
\end{eqnarray}
it is possible to  relate the measured classical intensities directly to the unknown tritter couplings, $g(z)$ and $G(z)$ or four-port splitting ratios and cancel the loss terms.

For the tritter we fitted a model to the classical measurements by assuming uniform coupling over an interaction region of length $L$, i.e. $g(0\leq z\leq L)L/\nu_p  = \bar{g} $ and  $G(0\leq z\leq L)L/\nu_p = \bar{G} $ with zero coupling elsewhere. Using maximum likelihood estimation~\cite{maxlike}, we varied  $\bar{g}$ and $\bar{G}$ in the transfer matrix $U$ to find the best fit to the three experimentally determined fractions $F^{1,2}_{1,2}$, $F^{1,3}_{1,3}$ and $F^{2,3}_{2,3}$. This involved minimizing the difference between the experimentally measured left hand side and theoretical right hand side of Eq. (\ref{Fractions}) for $F^{1,2}_{1,2}$, $F^{1,3}_{1,3}$ and $F^{2,3}_{2,3}$, weighted by the uncertainties in the fractions (found from the uncertainties of the measured intensities $N^k_j$) \cite{MLE}. This fit determined $\bar{g} \approx 0.81 \nu_p/L$ and $\bar{G} \approx 0.51 \nu_p/L$, indicating that waveguides one and two couple more strongly than waveguides one and three.  Using this fit we obtained the matrix of classical output intensities
\begin{equation}
|U_{tritter}|^2 = \left(
\begin{array}{ccc}
 0.37 & 0.41 & 0.23 \\
 0.41 & 0.19 & 0.41 \\
 0.23 & 0.41 & 0.37
\end{array}
\right).
\end{equation}
The ideal symmetric tritter has $|U_{i,j}|^2 = 1/3$, therefore we predict our device has a slightly asymmetric power splitting~\cite{Tritterbulkoptic}.

Similar to the tritter, we can use a maximum likelihood approach to fit $\eta$, the beamsplitter reflectivity, from (\ref{UMatrixFourPort}) using the following six fractions, $F^{1,2}_{1,2}$, $F^{1,3}_{1,3}$, $F^{1,4}_{1,4}$, $F^{2,3}_{2,3}$, $F^{2,4}_{2,4}$, and $F^{3,4}_{3,4}$. This fit determined $\eta= 0.377$.  Substituting this fit into Eq.~(\ref{UMatrixFourPort}) we find the matrix of classical output intensities for the four-port
\begin{equation}
|U_{4port}|^2 = \left(
\begin{array}{cccc}
 0.14 & 0.23 & 0.23 & 0.39 \\
 0.23 & 0.14 & 0.39 & 0.23 \\
 0.23 & 0.39 & 0.14 & 0.23 \\
 0.39 & 0.23 & 0.23 & 0.14
\end{array}
\right).
\end{equation}
Our four-port device differs from the ideal symmetric four-port, $|U_{i,j}|^2 = 0.25$  and has a non-uniform power splitting~\cite{Tritterbulkoptic}.This is due to the fact that the value we measure for $\eta = 0.377$ is not the ideal value of $\eta = 0.5$.

\section{Quantum characterisation}

We performed two photon characterisation on the devices using 804~nm photons produced using type-I spontaneous parametric down conversion (SPDC) source, as described in~\cite{Grahamoptex}. The setup used a Toptica 402~nm blue diode laser which was focused into a 1~mm thick BiBO crystal cut to give a $6\,^{\circ}$ opening cone angle. Down converted photons were produced at 804~nm in an output cone and then coupled into polarisation maintaining single mode fibres. These were then butt-coupled to the device under test using 127~$\mu$m spaced V-groove arrays. The outputs were then monitored using silicon avalanche photodiode detectors (SAPD) from Perkin Elmer. The 2-fold coincidences across all the output modes were measured simultaneously (in a 5~ns window) using a 4 channel time tagging unit. The arrival time of the photons  into the chips was varied as a means of continuously controlling the degree of indistinguishability between the two photons and thereby moving between classical propagation of light and quantum interference.

We detected signatures of two photon interference in the devices through the Hong-Ou-Mandel (HOM) effect~\cite{HOM}, whereby the probability of two photon coincidences are reduced (HOM dip) or enhanced (HOM peak)  due to quantum interference in the device. The degree of quantum interference is quantified using the visibility, $V^{k,l}_{i,j}$, defined for injecting a single photon into waveguides $i,j$ and detecting in waveguides $k,l$ as,
\begin{eqnarray}
V^{k,l}_{i,j} &=&  \frac{C^{k,l}_{i,j}-Q^{k,l}_{i,j}}{C^{k,l}_{i,j}},
\label{Visibilities} 
\end{eqnarray}
where, the quantum and classical coincidence probabilities are,
\begin{eqnarray}
Q^{k,l}_{i,j} &=& \frac{1}{1+\delta_{i,j}} | U_{i,k} U_{j,l} + U_{i,l} U_{j,k}|^{2}\\
C^{k,l}_{i,j} &=&  | U_{i,k} U_{j,l}|^{2} + |U_{i,l} U_{j,k}|^{2}.
\end{eqnarray}
We see that the quantum two photon coincidence probability $Q^{k,l}_{i,j}$ is strongly dependent on the relative phases of the elements of $U$. When these phases destructively interfere, two photon coincidences are reduced compared to the classical case, leading to a large positive visibility (a HOM dip). Whereas when these phases constructively interfere, two photon coincidences  are enhanced compared to the classical case, leading to a large negative visibility (a HOM peak).

\subsection{Visibilities in tritter}

\begin{figure}[H]
\centering
 \subfloat[]{\label{fig:VisibilityvsLengthSym}\includegraphics[width=0.48\textwidth]{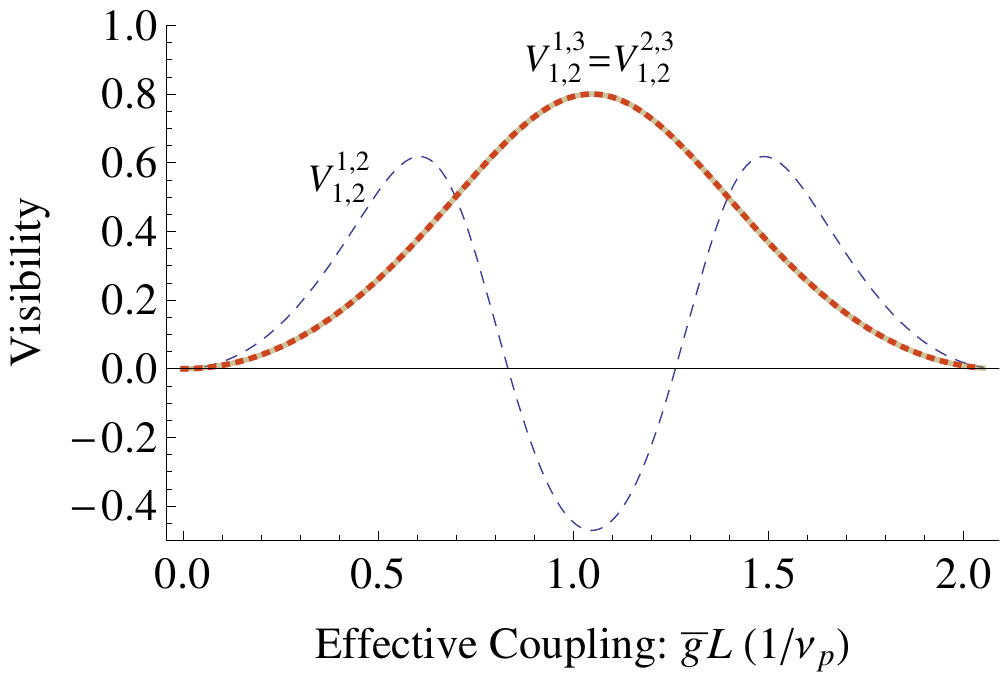}}
 \subfloat[]{\label{fig:VisibilityvsLengthNOTsym}\includegraphics[width=0.48\textwidth]{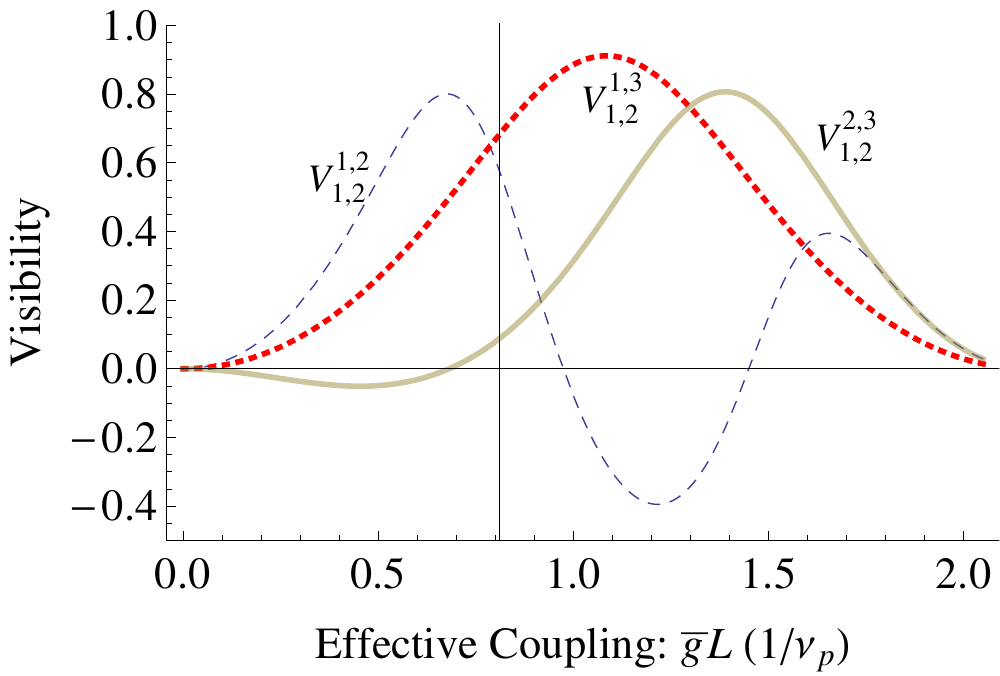}}\vspace {-4mm}  
\caption{ Tritter visibilities as a function of effective coupling, $\bar{g} L$. We assume uniform coupling over an interaction region of length $L$, $g(0\leq z\leq L)L/\nu_p  = \bar{g} $ and  $G(0\leq z\leq L)L/\nu_p = \bar{G} $ with zero coupling elsewhere. a) symmetric coupling region ($\bar{G}=\bar{g}$). b) Isosceles coupling region ($\bar{G}= 0.6234\bar{g}$).  Vertical line corresponds to the fit to our experimental tritter with  $\bar{g} L\approx 0.81 \nu_p$.
 }
 \label{fig:TritVvsLTheory}
\end{figure}

As the visibilities are strongly dependent on the relative phases of the matrix elements of $U$, the tritter visibilities depend on the length and  geometry of the coupling region as defined by the coupling functions $g(z)$ and $G(z)$ in Eq. (\ref{GeneralHam}). To illustrate the variety of tritter visibilities that can be observed in our 3D device, we plot the visibilities as a function of effective coupling $\bar{g} L$ in (Fig.~\ref{fig:TritVvsLTheory}) assuming uniform coupling over the interaction region. Figure~\ref{fig:VisibilityvsLengthSym} shows a symmetric tritter as gL is varied. Only at one coupling ( the symmetric 33/33/33 case) do we see equal visibilities of 50

\begin{figure}[h]
\centering
 \subfloat[]{\label{fig:Inputports1_2}\includegraphics[width=0.48\textwidth]{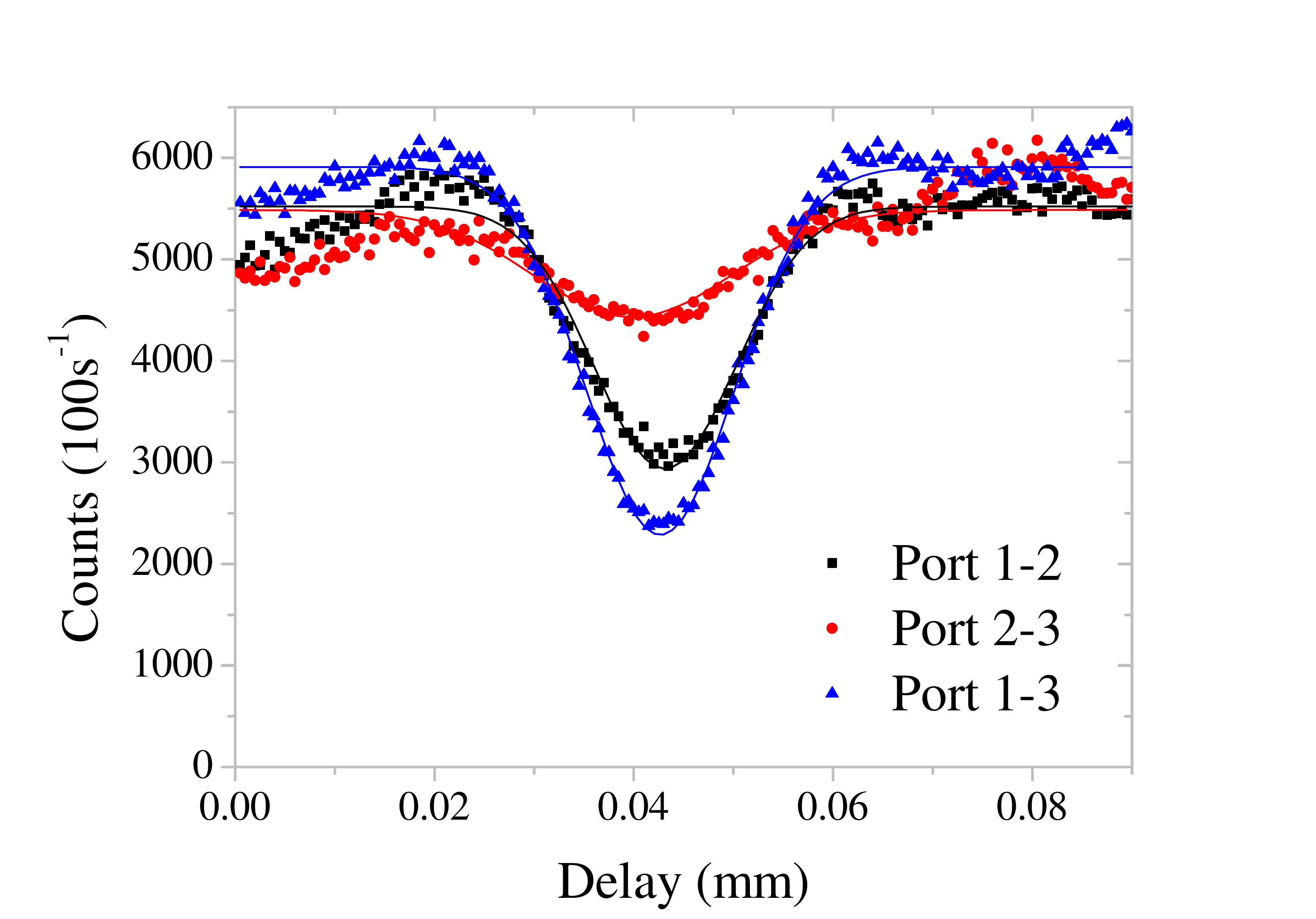}}
 \subfloat[]{\label{fig:Inputports2_3}\includegraphics[width=0.48\textwidth]{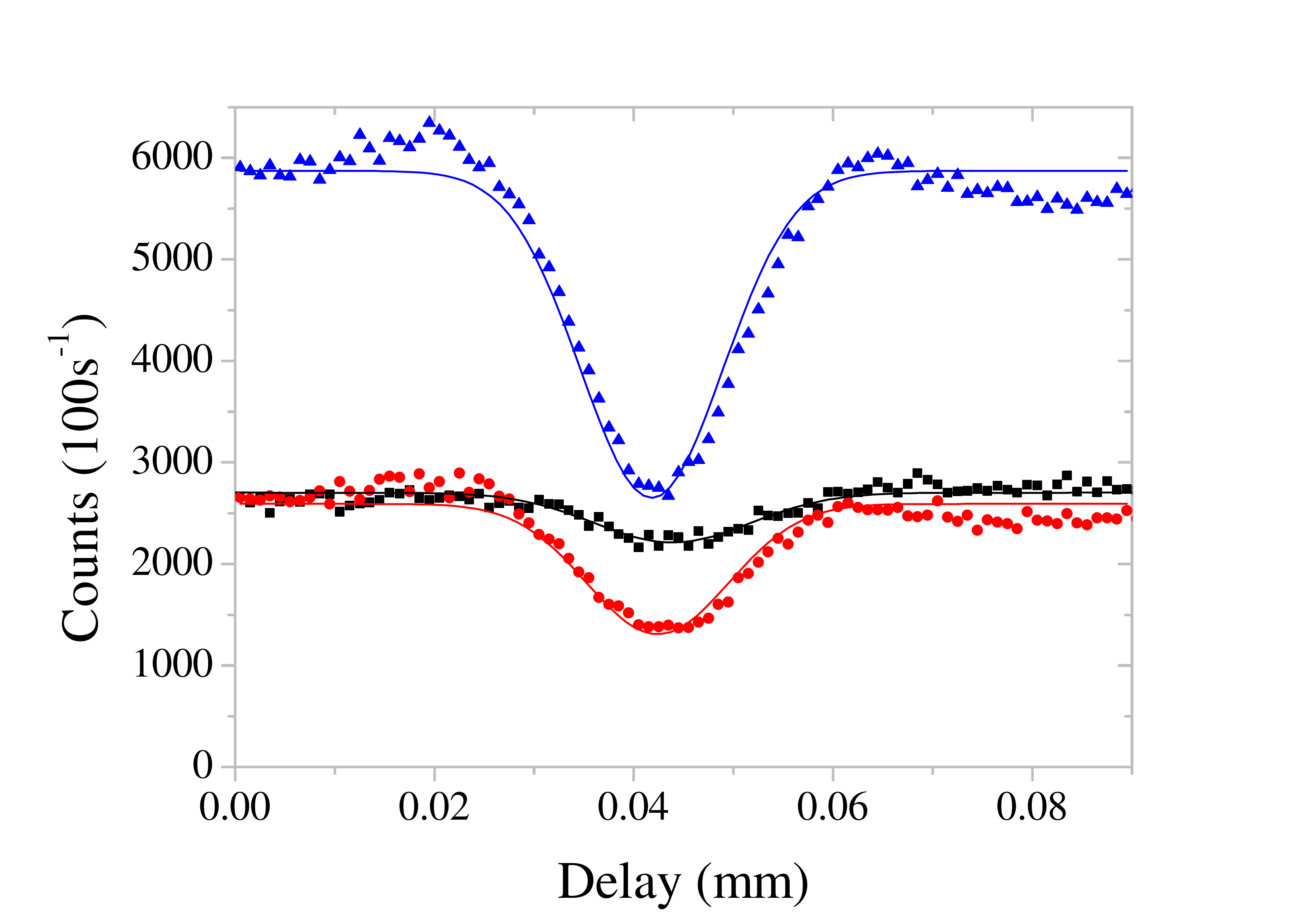}}\vspace {-4mm}  
 \subfloat[]{\label{fig:Inputports1_3}\includegraphics[width=0.48\textwidth]{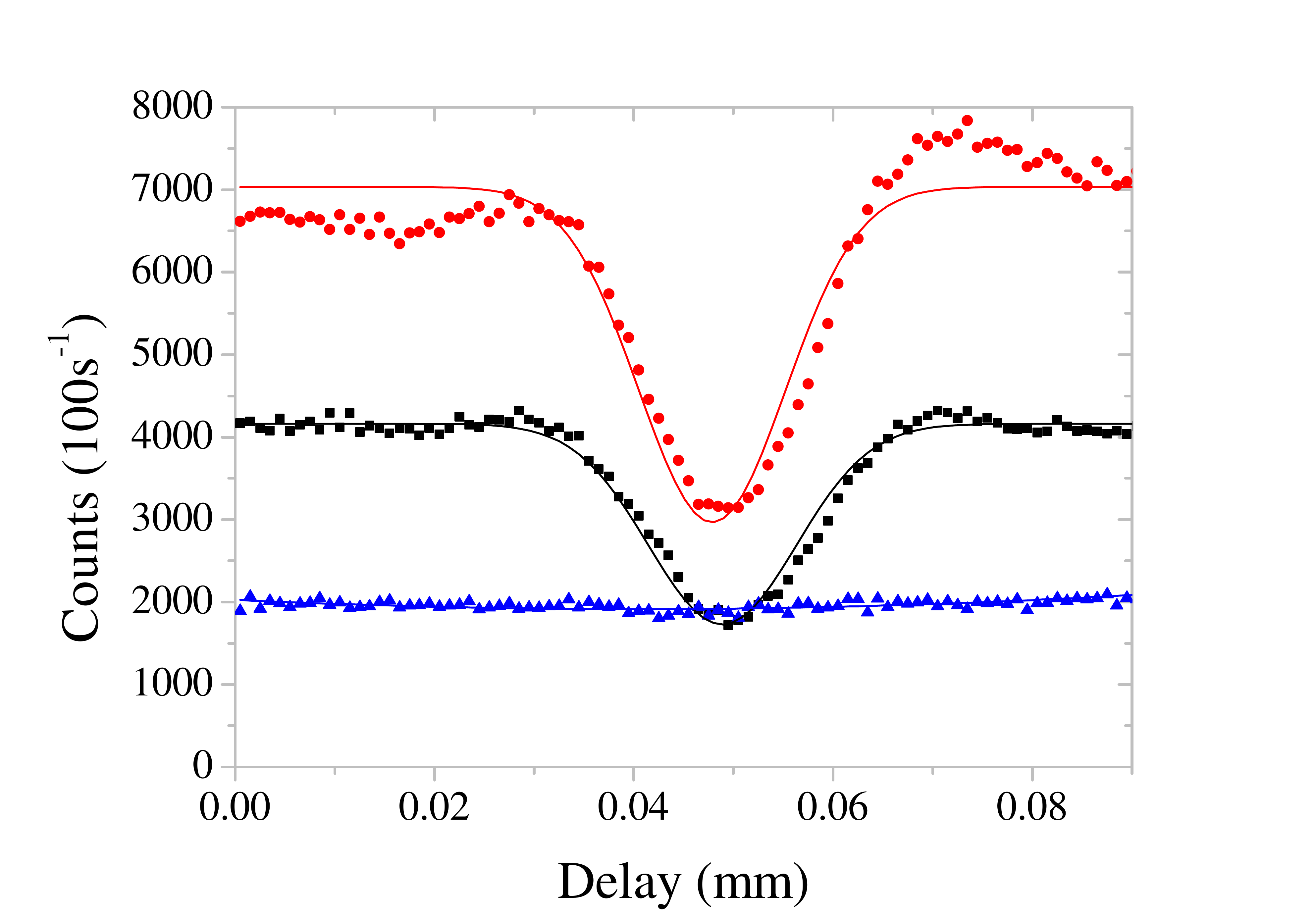}} 
\subfloat[]{\label{fig:tritvisTheory}\includegraphics[width=0.48\textwidth]{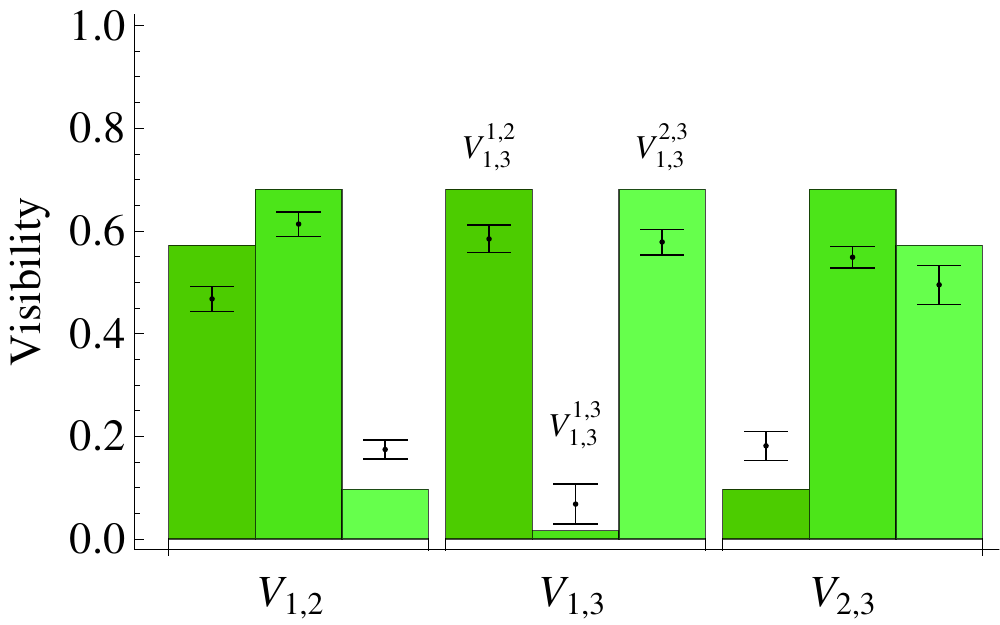}}\vspace {-4mm}
\caption{ Tritter HOM dips for (a) input ports 1 and 2, (b) for input ports 2 and 3 (c) for input ports 1 and 3. (d)  Comparison of measured tritter visibilities (error bars)  and theoretical predictions (columns) from the simple tritter model in section \ref{ClassicalSection}. }
 \label{fig:HOMdip}
\end{figure}

We measured nine HOM dips for the tritter (Fig.~\ref{fig:HOMdip}) by injecting two photons into the device at different delays and observing coincidences at the output ports. The  measured visibilities (Fig.~\ref{fig:tritvisTheory}) were found from the relative depth of each dip. We see a clear reduction in the visibilities observed in the cases of $V^{2,3}_{1,2}$, this corresponds to Fig.~\ref{fig:VisibilityvsLengthNOTsym}, we also see reductions in the visibilities for different input ports $ V^{1,3}_{1,3}$ and $V^{1,2}_{2,3}$.

In section \ref{ClassicalSection} we fitted a simple model of the tritter couplings $g(z)$ and $G(z)$ using  only classical intensity measurements of the device. As this fit uniquely determines $U$ we can predict the visibilities for this model using Eq.~(\ref{Visibilities}). Therefore we are able to predict, using only classical intensity measurements measurements, the quantum correlations for the device. This is a convenient tool, since typically in order to produce a device with the correct coupling ratios a range of device must be fabricated around this indended parameter. It is much more convenient to perform a range of intensity measurements prior to any quantum measurements to confirm the device has the desired functioning. The predicted visibilities are compared to the experimental visibilities in Fig.~\ref{fig:tritvisTheory}, where we see a qualitative match, issues which are causing the minor mismatch between this prediction and the experimentally observed values are the assumptions uppon which the intensity dependent measurement is based. The model assumes a constant input intensity which can fluctuate during the course of the measurement and input output facet losses which may change also. Furthermore the model assumes the interaction region is constant in z but input and out tapering regions do contribute to the overall coupling which was not accounted for. In addition since the maximum likelihood measurment is a numerical simulation it has it's own inherent errors~\cite{maxlike,mqubit}.

\subsection{4-port}
\begin{figure}[H]
\centering
 \subfloat[]{\label{fig:4port_input1-2}\includegraphics[width=0.48\textwidth]{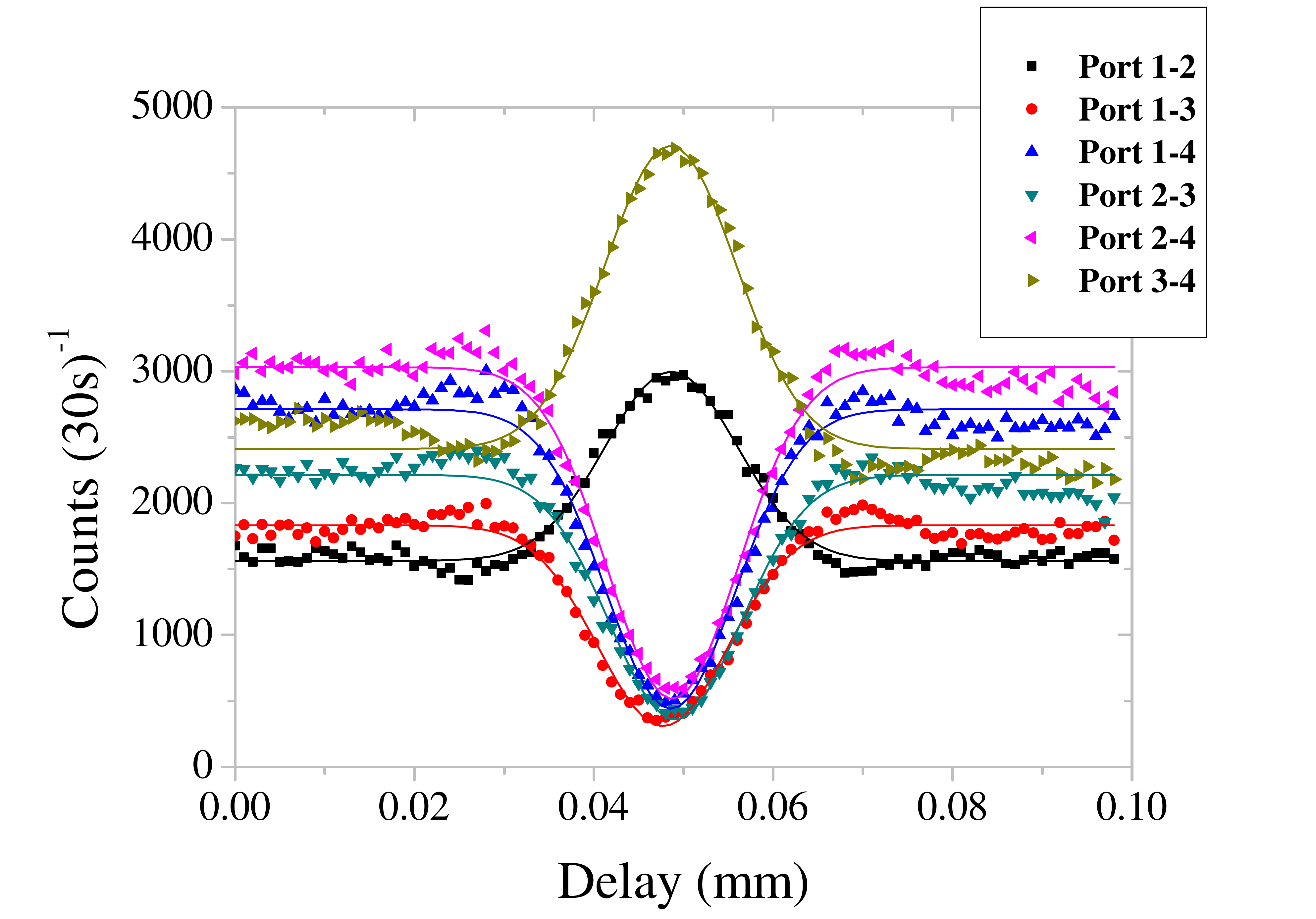}}
 \subfloat[]{\label{fig:4port_input1-3}\includegraphics[width=0.48\textwidth]{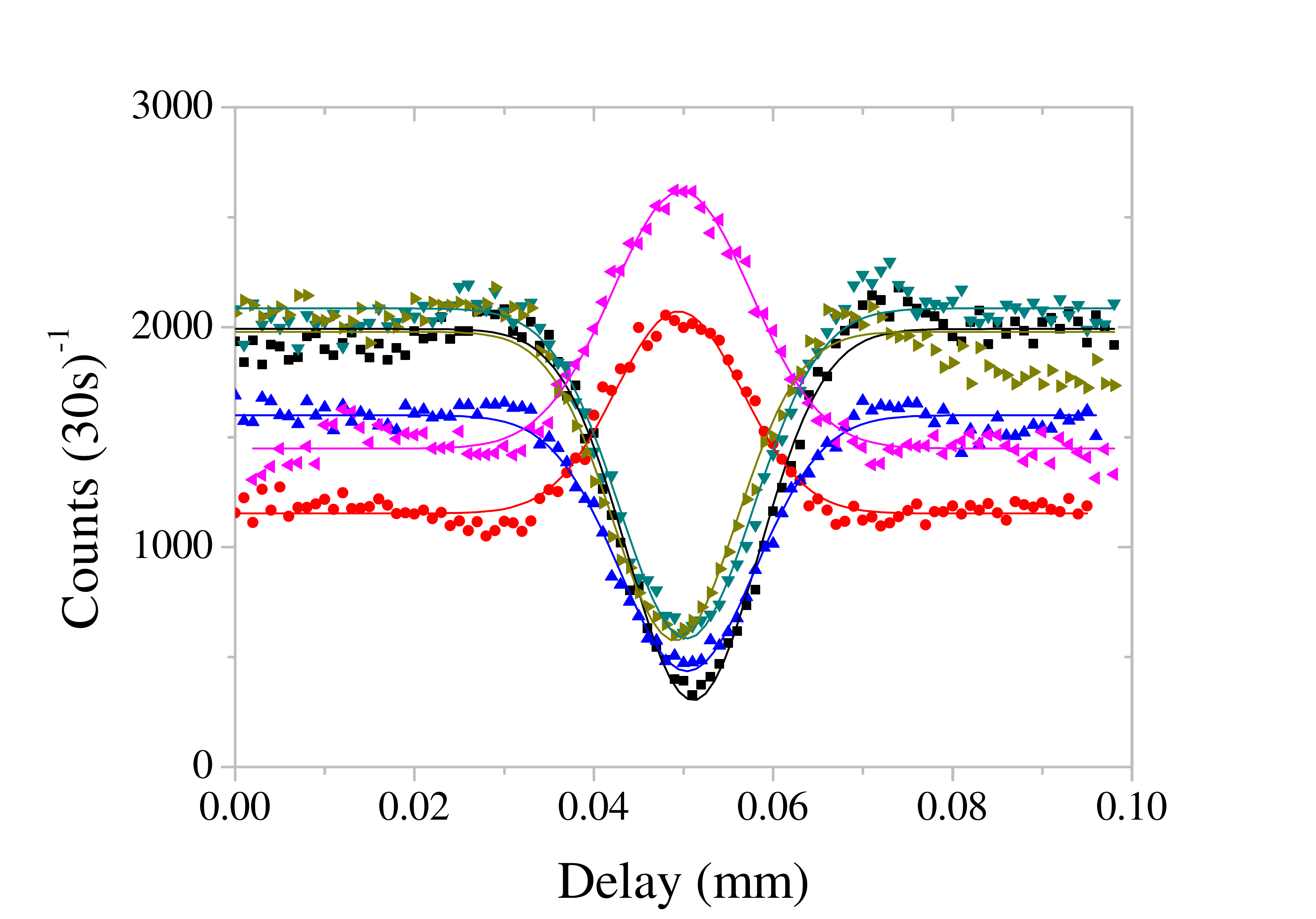}}\vspace {-4mm}  
 \subfloat[]{\label{fig:4port_input1-4}\includegraphics[width=0.48\textwidth]{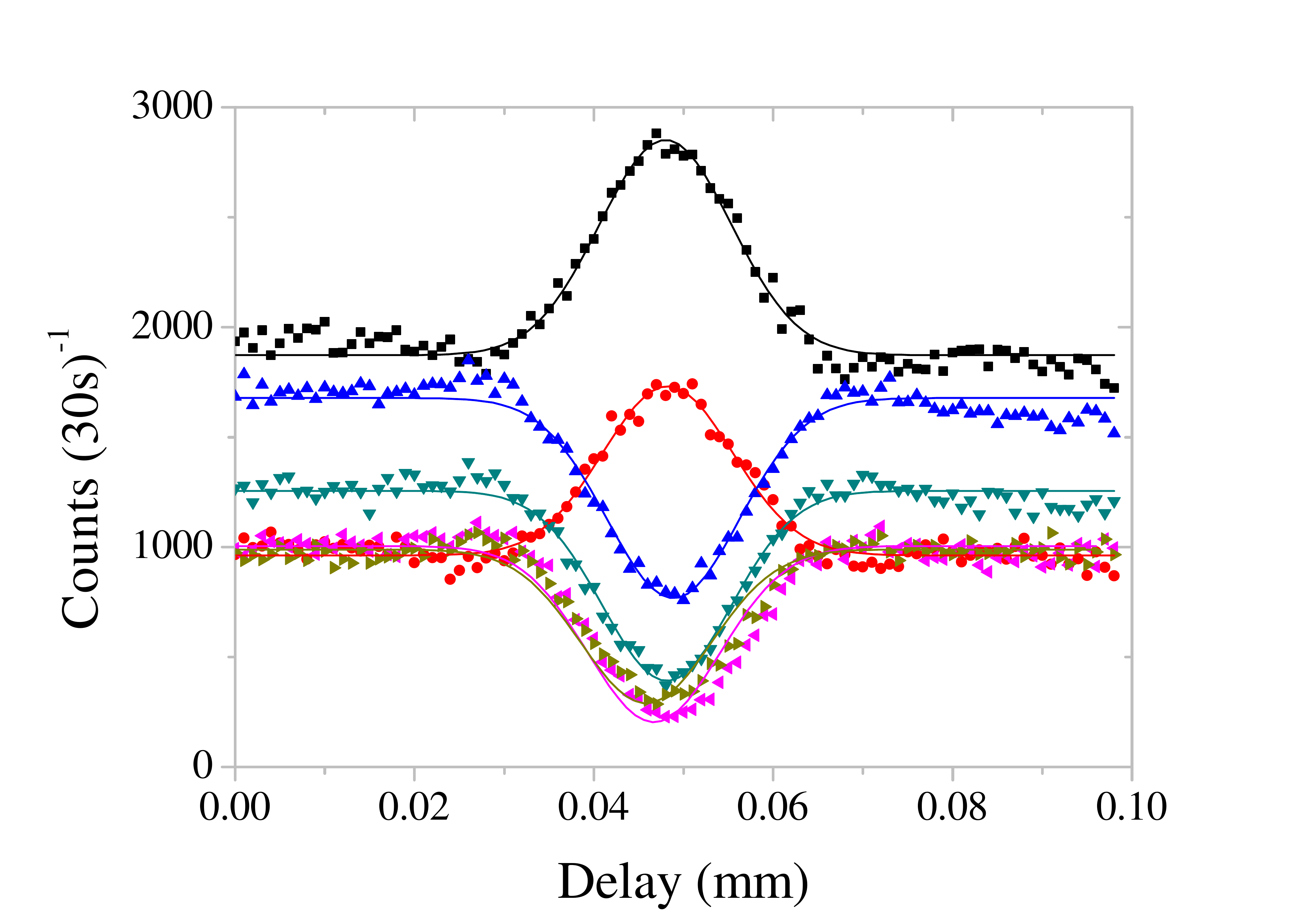}}
 \subfloat[]{\label{fig:4port_input2-3}\includegraphics[width=0.48\textwidth]{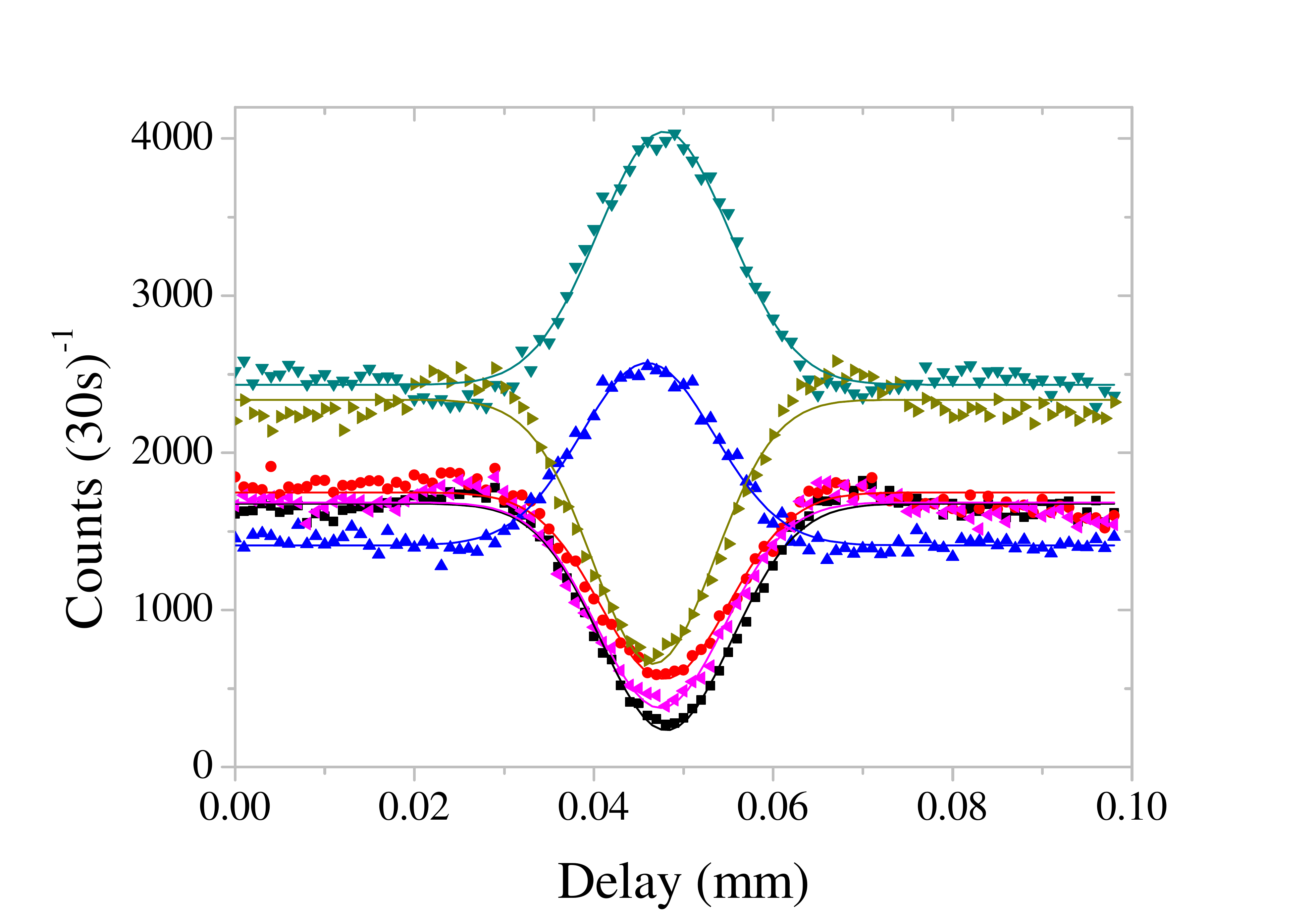}}\vspace {-4mm}
 \subfloat[]{\label{fig:4port_input2-4}\includegraphics[width=0.48\textwidth]{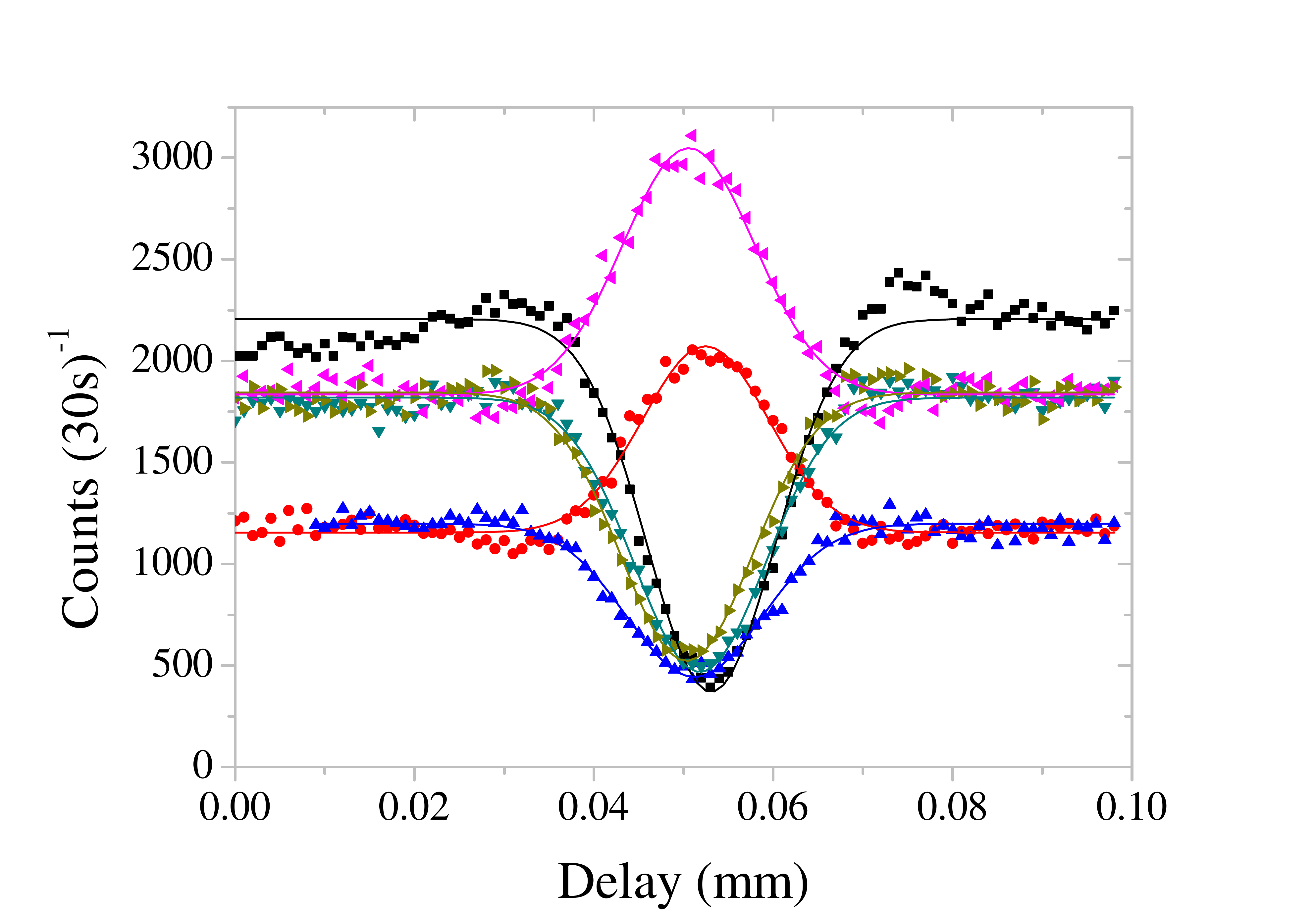}}
 \subfloat[]{\label{fig:4port_input3-4}\includegraphics[width=0.48\textwidth]{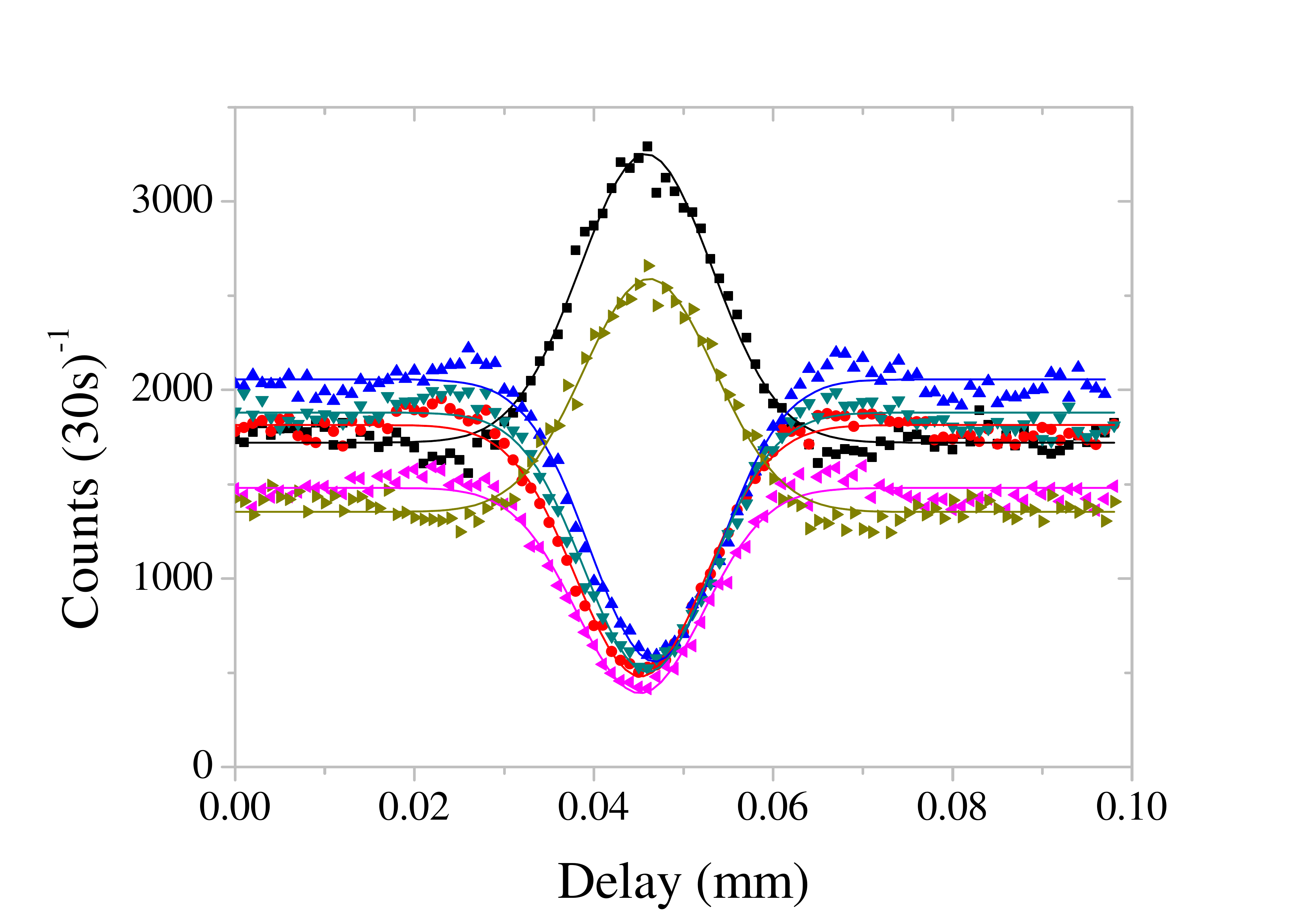}}\vspace {-1mm}
\subfloat[]{\label{fig:4port_theory}\includegraphics[width=0.88\textwidth]{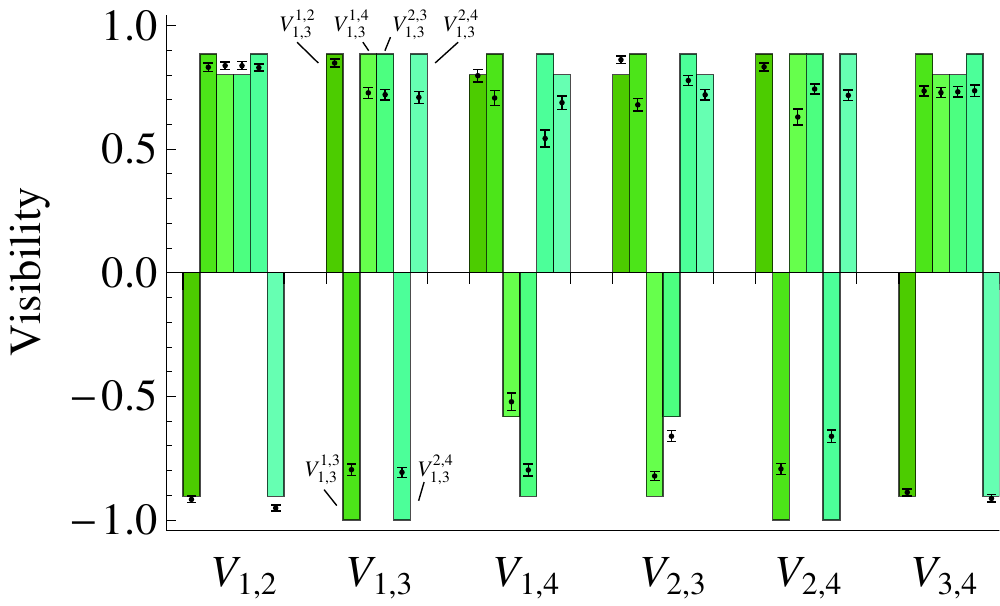}}
\caption{ Four-port HOM dips for (a) input ports 1 and 2, (b) for input ports 1 and 3 (c) for input ports 1 and 4 (c) for input ports 2 and 3 (d) for input ports 2 and 4 (e) for input ports 3 and 4.  (f) Experimental four-port visibility matrix determined from the HOM dips. (g) Comparison of measured four port visibilities (error bars) and theoretical predictions (columns) from the four-port model.}
 \label{fig:4_port_HOMdip}
\end{figure}

We also measured HOM interference for all input and output combinations in the 4-port device. These measurements are shown in Fig.~\ref{fig:4port_input1-2}--\ref{fig:4port_input3-4}. We see a variety of different magnitudes and signs of the visibilities for the various input/output combinations. This is due to the  quantum term  $Q^{k,l}_{i,j}$ in Eq.~(\ref{Visibilities}), which  exhibits constructive or destructive  interference, depending on the relative phases of the elements of $U$. The fact that all outputs had a visibility greater than $50 \%$ (or less than ($-50\%$)) indicates that the output states are strongly non-classical.

We now compare the experimental visibilities to those predicted from the 4-port model in section \ref{ClassicalSection}. In contrast to the tritter the classical intensity measurements in the 4-port are insufficient to determine the phase $\phi$ due to the additional path lengths of waveguides two and three.  Therefore after fitting $\eta= 0.377$ from the classical intensity measurements we used visibility measurements from a single input, namely inputs two and three, to fit $\phi$. Using maximum likelihood estimation we fitted $\phi$ from the six HOM dips in Fig.~\ref{fig:4port_input2-3}. This fit determined $\phi =0.07 \pi$. After substituting the determined $\eta$ and $\phi$ into the transfer matrix Eq.~(\ref{UMatrixFourPort}) we can use (\ref{Visibilities}) to predict the other $30$ visibilities in  Fig.~\ref{fig:4_port_HOMdip}. The predicted visibilities are compared to the measured visibilities (see Fig.~\ref{fig:4port_theory}) and display a qualitative match, correctly identifying the sign and magnitude of each visibility. We note that using other input combinations to determine $\phi$ produced similar predictions.

\section{Conclusion}
Two photon correlation measurements were performed on integrated 3 and 4 port devices. The visibilities observed for our tritter compare extremely well with bulk optic demonstrations where visibilities, for only one pair of input ports, $V_{1,2}^{1,2} \sim 50\%, V_{1,2}^{1,3} \sim 48\%$  and $V_{1,2}^{2,3} \sim 26\%$ are observed~\cite{Tritterbulkoptic}. In addition our tritter displays greater symmetry than a fibre optic based example where visibilities, observed only for a single pair of input ports $V_{1,2}^{1,2} \sim 50\%, V_{1,2}^{1,3} \sim 30\%$  and $V_{1,2}^{2,3} \sim 25\%$~\cite{Tritterfibre}. Our four port has shown a uniform output very similar to a bulk optic example, where visibilities of $V_{1,2}^{1,2} \sim 75\%, V_{1,2}^{1,3} \sim 66\%, V_{1,2}^{1,4} \sim - 66\%, V_{1,2}^{2,3} \sim - 80\%, V_{1,2}^{2,4} \sim 75\%$  and $V_{1,2}^{3,4} \sim 66\%$ were observed~\cite{Tritterbulkoptic} which out performs a multimode interference device described in~\cite{MMI}. This demonstrates the applicability of laser written photonic circuits for some quantum logic applications where a 3D capability is advantageous for circuit size or complexity requirements.

\section*{Acknowledgments} 
This research was conducted in part by the Australian Research Council Centre of Excellence for Ultrahigh bandwidth Devices for Optical Systems (project number CE110001018).
This work was performed in part at the Macquarie University node of the Australian National Fabrication  Facility. A company established under the National Collaborative Research  Infrastructure Strategy to  provide nano and microfabrication facilities for Australia¡'s  researchers. 
 GDM acknowledges support from the Australian Academy of Science's International Science Linkages scheme and the Marie Curie International Incoming Fellowship scheme.
We acknowledge the generosity of Prof. Gabriel Molina-Terriza and Dr Mathieu Juan in lending equipment and Dr. Douglas J. Little for offering valuable insights and observations.

\end{document}